# Characterization of a Differential Radio-Frequency Single-Electron Transistor

J.F. Schneiderman*, P. Delsing[†,‡], M.D. Shaw*, H.M. Bozler*, and P.M. Echternach[‡]

[‡]*Jet Propulsion Laboratory, California Institute of Technology, Pasadena, CA 91109, USA*
*University of Southern California, Dept of Physics and Astronomy, Los Angeles, CA 90089-0484, USA*
[†]*Chalmers University of Technology, Microtechnology and Nanoscience, MC2, 412 96 Göteborg, Sweden*

**Abstract.** We have fabricated and characterized a new type of electrometer that couples two parallel single-electron transistors (SETs) to a radio-frequency tank circuit for use as a differential RF-SET. We demonstrate operation of this device in summing, differential, and single-SET operation modes, and use it to measure a Coulomb staircase from a differential single Cooper-pair box. In differential mode, the device is sensitive to uncorrelated input signals while screening out correlated ones.

The single-electron transistor[1,2] (SET) is widely used as an extremely sensitive electrometer, having found a niche in a number of applications involving ultra-sensitive measurements. Unfortunately, background charge fluctuations[3] have been a problem plaguing single electron devices since their inception. Such charges are thought to be a primary cause of the 1/f noise that limits the resolution of precision charge measurements. Over time, a number of important technological improvements have been made to the SET, such as radio-frequency operation[4] (RF-SET). The RF-SET is operated at sufficiently high frequencies for the 1/f noise level to be substantially lowered.

In many applications, it is desirable to measure the difference in charge between two parts of the same object. For example, Buehler et al. recently used two independent RF-SETs to measure the motion of charge between two phosphorous dots embedded in silicon[5]. For this measurement, independent resonant circuits were used for each SET, and the readout signal was subtracted after RF detection. Furthermore, one proposed variety of superconducting charge qubit is the differential single Cooper-pair box (DSCB)[6,7], which requires measuring the difference in charge between two islands separated by a Josephson junction.

In this letter, we report on the experimental demonstration of a differential RF-SET (DRFSET) consisting of two parallel SETs acting as the dissipative element of a single resonant LC circuit (see Figure 1). Each SET has a separate transfer function which reflects the change in its conductance as a voltage is applied to its respective gate. To achieve differential readout, the SETs are biased so that their responses are of opposite sign. In this mode, when a fluctuation in charge couples to both SET islands, the conductance change of one island will be opposite in sign to that of the other, leading to a cancellation of the change in the conductance of the parallel SETs. In other words, the DRFSET is insensitive to collective fluctuations in charge affecting both SET islands together. This mode of operation also allows for a strong differential readout since increasing the charge coupled to one SET while decreasing the charge coupled to the other will produce a correlated and reinforced change in conductance of the parallel SETs, and hence a large change in the dissipation of the LC tank circuit for readout.

Other modes of DRFSET operation are also possible, depending on where each SET is biased on its transfer function (see Figure 2). The simplest mode of operation is to bias one SET at an insensitive part of its transfer function and the other at its highest sensitivity. This "single-SET" mode is useful for characterization of the separate SETs. Finally, one can operate the DRFSET in a "common" or summing mode which is closer to the usual RF-SET setup.

Summing mode is made possible by biasing each SET island so that the signs of their respective responses are the same. Both SETs will then respond in the same way to an overall change in charge on their respective islands.

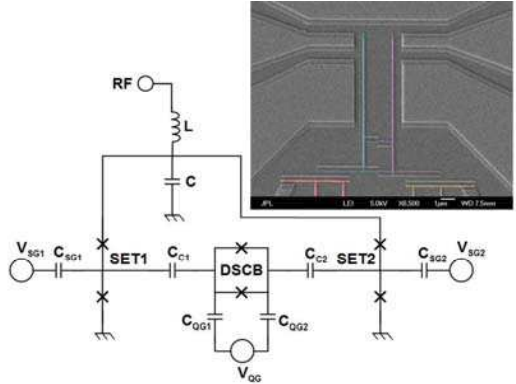

**FIGURE 1.** Circuit diagram showing the DSCB with DRFSET readout. $C_{SG1}$, $C_{SG2}$, $C_{QG1}$, $C_{QG2}$, $C_{C1}$, and $C_{C2}$ are the SET gate, qubit gate, and coupling capacitances for SET1 and SET2, respectively. L and C are the inductance and capacitance of the tank circuit. Inset: SEM image of the pictured circuit. The red and yellow features are the two SETs, while the cyan and violet features form the DSCB.

As mentioned above, an application of the DRFSET of particular interest for quantum computing is a charge readout for the differential single Cooper-pair box (DSCB). The DSCB qubit consists of two islands coupled by a small tunnel junction, with the relevant quantum states being the difference in charge between the two islands. Even though the potential of the islands may fluctuate with offset charges, the difference in charge between them should be relatively unaffected unless the fluctuating charge is very close to the DSCB itself. The DRFSET becomes an ideal readout device for the DSCB by coupling each of the DRFSET islands to a separate DSCB island.

We have achieved operation of a differential RF-SET and used it as a differential electrometer to measure a Coulomb Staircase from a DSCB. The inset of Figure 1 shows an SEM picture of a DSCB with its left and right islands coupled to separate islands of a DRFSET. The samples were fabricated using a standard double-angle shadow mask evaporation technique[2]. Each island of the DSCB has a gate electrode which is used to push and pull charges from one island to the other, while each SET has its own tuning gate used to bias it at different operating points. Operation in the RF mode is achieved by sending an RF signal to an LC tank circuit for which the parallel SETs act as the dissipative element[4].

We have measured the voltage of the reflected signal as a function of both gate voltages with the SETs biased at the double-Josephson-quasiparticle (dJQP) peak[8,9]. Figure 2 shows the results of such a measurement, and indicates gate voltage values appropriate for each mode of operation. Regions where the slope of the curve is large represent operating points of high SET sensitivity. At point **a**, for example, the reflected signal is essentially independent of the voltage on gate 1, but depends strongly on gate voltage 2, and thus represents a point where only SET2 is sensitive (single-SET mode). Similarly, at point **b** only SET1 is sensitive. At points **c** (summing mode) both SETs are sensitive and biased on the same slope, so an increase in both gate voltages yields a net sum change in the reflected RF signal.

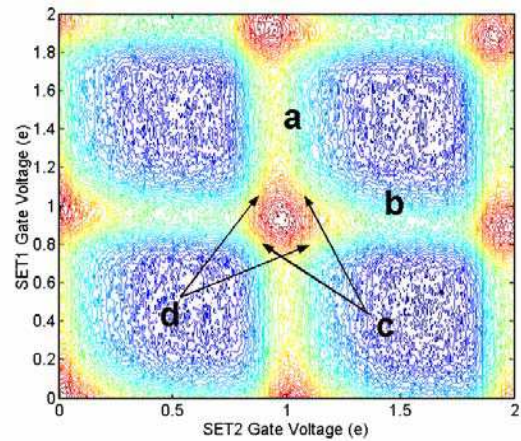

**FIGURE 2.** Coefficient of reflected power as a function of both SET gate voltages. Four modes of operation are shown: **a**, single-SET mode, SET1 insensitive; **b**, single-SET mode, SET2 insensitive; **c**, summing mode; **d**, differential mode.

Finally, at points **d** (differential mode), where the SETs are biased on opposite slopes, an increase in both gate voltages results in an opposing change in the conductance of each SET, tending to cancel out the change in reflected power from the overall device. However, a change in each gate voltage in opposite directions will yield a reinforced change in the overall reflected power. This makes the DRFSET an ideal device for reading out the charge state of a DSCB.

To demonstrate operation of the DRFSET, we applied small amplitude ($0.01e_{PP}$) low-frequency signals to each of the SET gates, and a third signal to both qubit gates. Each signal had a unique and coprime frequency to avoid self-mixing or mixing with harmonic and subharmonic modes. We applied 9- and 11-Hz signals to SET1 and SET2 respectively, and a 13-Hz "common-mode" signal to the gates of the DSCB, which couples to the SET gates with equal strength via the cross-capacitances. By varying the DC offset levels of the 9- and 11-Hz signals, we could vary the operating point of the DRFSET, changing between summing, differential, and single-SET modes. We examined the reflected signal from the DRFSET using a spectrum analyzer, as shown in Figure 3. In Figures 3a and 3b, the SETs are biased in single-SET mode as described above.

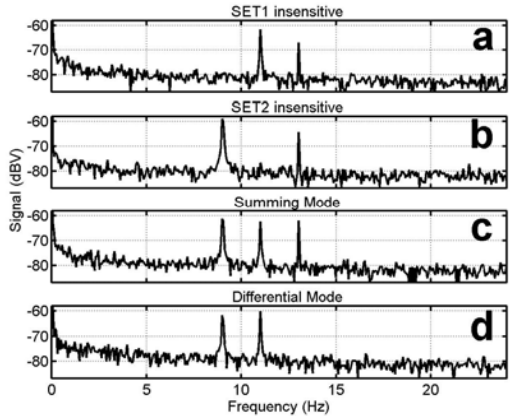

**FIGURE 3.** DRFSET readout spectra. Graphs a-d correspond to the modes of operation described in Figure 2.

Figure 3c shows the operation of the DRFSET in the summing mode, while Figure 3d shows results for operation in differential mode. Observe the peak corresponding to the 13-Hz common mode signal is quite pronounced while in summing mode, whereas operation in the differential mode causes this signal to be suppressed below the 1/f noise floor. This demonstrates that a DRFSET operated in differential mode is insensitive to a common mode signal, and thus has the advantage of being immune to correlated noise.

In a similar fashion, we examined noise spectra at a variety of different SET operating points. At high frequencies (tens of kHz and above), the noise floor remained unchanged even when comparing areas of high SET sensitivity to areas of total insensitivity. This result can be explained by the assumption that the noise floor is dominated by amplifier noise above a few kHz. However, at frequencies below a few hundred Hz, noise levels followed the gain of the SETs, irrespective of readout mode. We ensured the DRFSET was insensitive to correlated input signals not only for the select frequency used for the "common mode" signal described above, but for all frequencies up to several hundred Hz. We applied a common white noise signal to both qubit gates and measured spectra in the differential and summing readout modes. The differential readout mode showed a reduction in the correlated noise by 9.8±0.2 dB. The fact that the level of the 1/f noise floor does not change between summing and differential modes implies there is no correlation in the noise felt by the two SETs. This in turn indicates that most of the 1/f noise results from charge fluctuators which are much closer to either SET than the distance between each SET (5 μm). This result is compatible with those of Zorin et al.[3], which indicate a 15% correlation of noise for SETs separated by 0.2 μm.

As a further demonstration of the DRFSET, we have measured a Coulomb staircase[10] from a differential single Cooper-pair box with the DRFSET. By applying a voltage ramp to one side of the DSCB and a similar ramp of opposite sign to the other, charge was transferred from one island of the DSCB to the other. This transfer of charge was then measured by the DRFSET operating in differential mode.

In conclusion, we have experimentally demonstrated a differential radio-frequency single electron transistor in its four modes of operation. Differential readout was verified by measuring a Coulomb staircase from a differential single Cooper-pair box. At low frequencies, the 1/f noise floor was shown to be independent of the operating mode, implying a lack of 1/f noise correlation between the two SET islands situated 5 μm apart.

We thank Rich E. Muller for the electron-beam lithography. Research was performed at the Jet Propulsion Laboratory, California Institute of Technology, under a contract with the National Aeronautics and Space Administration. We acknowledge support by the National Security Agency and the Advanced Research and Development Activity.